\documentstyle[prd,aps,amsmath,epsf,graphicx]{revtex}

\DeclareFontFamily{OT1}{rsfs10}{}
\DeclareFontShape{OT1}{rsfs10}{m}{n}{ <-> rsfs10 }{}
\DeclareMathAlphabet{\mathscript}{OT1}{rsfs10}{m}{n}

\newcommand{\be}{\begin{equation}}
\newcommand{\ee}{\end{equation}}

\newcommand{\bea}{\begin{eqnarray}}
\newcommand{\eea}{\end{eqnarray}}
\newcommand{\ba}{\begin{array}}
\newcommand{\ea}{\end{array}}

\newcommand{\ns}{\normalsize}

\def\a{\alpha}

\def\e{\epsilon}

\def\k{\kappa}

\def\cV{{\cal V}}

\begin{document}

\begin{titlepage}

\title{
\hfill{\ns SUSX-TH/02-009\\}
\hfill{\ns hep-th/0309096\\[2cm]}
{\Large Gauge Five Brane Moduli In Four-Dimensional Heterotic Models.}
\\[1cm]}
\setcounter{footnote}{0}
\author{{\ns\large
  James Gray$^1$\footnote{email: J.A.Gray2@newcastle.ac.uk}~ and
\setcounter{footnote}{2}
 Andr\'e Lukas$^2$\footnote{email: a.lukas@sussex.ac.uk} \\[1cm]}
      {\ns $^1$Department of Mathematics and Statistics, University of Newcastle upon Tyne,}\\
      {\ns Newcastle upon Tyne WEI 7RU,
      United Kingdom}\\[0.2cm]
      {\ns $^2$Department of Physics and Astronomy, University of Sussex}\\
      {\ns Falmer, Brighton BN1 9QH, United Kingdom}\\[0.2cm]}


\maketitle

\vspace{2cm}

\begin{abstract}
We present a K\"{a}hler potential for four dimensional 
heterotic M-theory which includes moduli describing a gauge five brane living on one of the orbifold 
fixed 
planes. This result can also be thought of as describing compactifications of either of the weakly 
coupled heterotic strings in the presence of a gauge five brane. This is the first example of a 
K\"{a}hler potential in these theories which includes moduli 
describing background gauge field configurations. Our results are valid when the solitons width is 
much smaller than the size scale of the Calabi-Yau threefold and can be used to provide a more 
complete description of some moving brane scenarios. 
We point out that, in general, it is not consistent to truncate away the gauge five brane moduli in 
a simple manner.
\end{abstract}

\thispagestyle{empty}

\end{titlepage}

\renewcommand{\thefootnote}{\arabic{footnote}}


\section{Introduction}
\label{intro}

Heterotic M-theory \cite{Witten:1996mz,Lukas:1998fg}, the compactification of the Ho\v{r}ava-Witten 
strongly coupled limit of the $E_8 \times E_8$ heterotic string \cite{Horava:1996qa,Horava:1996ma} on 
a manifold of $SU(3)$ holonomy, is one of the most promising corners of the M-theory moduli space 
studied to date from a phenomenological point of view. The theory combines phenomenological 
successes of its weekly coupled counterpart \cite{Green:mn} with a natural mechanism for obtaining 
the correct strength of gravitational interactions, through a kind of 'large extra dimensions' 
mechanism \cite{Witten:1996mz,Banks:1996ss}.

The vacuums associated with heterotic M-theory, which were presented in 
\cite{Witten:1996mz,Lukas:1998fg,Lukas:1999yy},  are non-trivial domain wall solutions with a 
warping in the 
direction of the bulk and, more importantly from the point of view of this paper, gauge field 
expectation values on at least one of the fixed planes. This gauge field background is taken to live 
entirely within the Calabi-Yau three-fold (i.e. it is not allowed to depend on the four external 
directions and is taken to be zero when its index is external) in order to maintain four dimensional 
Poincar\'{e} invariance. It is this vacuum which has been used in the study of four dimensional 
phenomenology and modulus evolution. Due to the models many successes from a particle 
physics standpoint extensive studies have been made of the moduli evolution about this vacuum, the 
basic solutions being provided in \cite{Brandle:2000qp}. However, to the authors knowledge, 
no one has included any of the moduli describing the 
background gauge field configuration in obtaining such cosmological solutions. The reason for this 
is quite simple - the relevant 
kinetic terms are not known (although some other information about these moduli has been obtained in 
\cite{He:2003tj,Buchbinder:2002ji,Buchbinder:2002ic,Buchbinder:2002pr,Buchbinder:wz}). The reason 
for this lack of a four dimensional theory is quite simple to understand. The most straightforward 
way to calculate such kinetic terms would be to start with a background solution which described a 
Calabi-Yau compactification of the theory, including the sections of the gauge bundles living on the 
fixed planes. One would then take the integration constants in this solution, promote them to be 
four dimensional fields and plug the resulting configuration into the higher dimensional action. 
Integrating out the internal dimensions would then naively result in the desired terms in the four 
dimensional effective action. However, such exact solutions on a compact Calabi-Yau threefold are 
not known rendering this calculation impossible.

The advent of some recent scenarios based upon moving and colliding M5 branes 
\cite{Khoury:2001wf,Bastero-Gil:2002hs} has made the need to include some of the gauge bundle 
moduli in our cosmological analysis even more pressing. The M5 branes concerned can be included in 
the vacuum solution without breaking $N=1$ supersymmetry if they are oriented parallel to the fixed 
planes in the bulk with two of their world volume directions wrapping a holomorphic curve within the 
Calabi-Yau \cite{Witten:1996mz,Lukas:1998hk}. The scenarios mentioned above are based upon the 
position modulus of an M5 brane evolving in such a way that 
the object collides with an orbifold fixed plane 
\cite{Derendinger:2001gy,Copeland:2001zp,Copeland:2002fv}. However it is not the case that during 
such a collision the M5 brane disappears with nothing else in the situation changing - this would 
result in an inconsistency in the cohomology condition, for example, which essentially says that the 
charges on the fixed planes and M5 branes should sum to zero. Various considerations, including 
study of the cohomology condition and examination of extra light states which appear at collision, 
lead us to believe that one thing that might happen on collision is a so called small instanton 
transition \cite{Witten:1996gx,Ganor:1996mu}. Here, during collision, the M5 brane disappears and is 
replaced with a gauge five-brane living on the relevant orbifold fixed plane. A gauge five brane is a 
solitonic object made completely out of low energy fields - including gauge fields 
\cite{Strominger:et}. The object appears with fundamental length scale width just after the 
collision and then could spread out with 
time to become more diffuse. The moduli which describe 
the gauge five brane are examples of gauge bundle moduli. The easiest way to understand this is to 
observe that the soliton is essentially a Yang-Mills instanton with various gravitational field 
dressings. Once the gauge field core of the object is known the dressing can be determined 
completely in terms of this (up to certain discrete choices which are available) and so the moduli 
describing the gauge five brane are simply the moduli of the Yang-Mills instanton - i.e. moduli 
associated with gauge bundle on that fixed plane.

As we shall see, bundle moduli cannot in general be consistently truncated off and, therefore, represent 
an essential part of heterotic low-energy effective theories which has been widely neglected so far. 
In particular, they must be included in cosmological scenarios, such as those mentioned above, where the 
gauge bundle may evolve in time. For example, one would like to understand whether or not the gauge five 
brane {\it does} spread out after a small instanton 
transition. A prerequisite for such an investigation would be knowledge of the kinetic terms for the
 appropriate moduli.

Given this situation in this paper we present a calculation of the effective four dimensional theory 
which 
describes the centred moduli space of the gauge five-brane (neglecting non-perturbative potentials). 
This theory contains, for instance, the size modulus for the soliton mentioned above. To our 
knowledge this is the first example of a K\"{a}hler potential which includes gauge bundle moduli 
describing the background configuration of gauge fields in Heterotic M-theory. 

In obtaining this four dimensional action we circumvent the problem of obtaining an explicit 
background solution with which to work by realising that the gauge five brane, at least when its 
width is small with respect to the curvature scale of the Calabi-Yau, is in some sense a very 
localised object. In such a regime the five-brane does not, outside of its core, probe the 
directions transverse to it to any significant degree \cite{Witten:1996gx}. In particular, in some 
senses, it does not know if the transverse space is compact or asymptotically flat. The idea then 
is to construct an 
approximate solution for the gauge and gravitational fields which is only valid close to the gauge 
five-brane (in a manner to be made explicit later). One then has to see if the effective action can 
be reliably calculated with only this limited information - i.e. can we calculate the effective 
theory describing the object without knowing what happens far from its world volume in the 
transverse 
space. We find that the answer is, as one might expect physically, in the affirmative.

Although we will be working in the heterotic M-theory set up here it should be stressed that similar 
configurations to the gauge five brane appear in many other phenomenologically viable 
compactifications of string theory. Many of the comments made above would equally well apply to 
these cases and one would expect the method we present to be viable there as well. For example one 
could consider a situation in type I where we have an instanton based configuration living on a 
stack of $D_p$ branes. Such a configuration could have been created by the collision of a $D_{p-4}$ 
brane \cite{Johnson:gi}. We would like to stress that these solitonic objects, as well as the ones we 
consider 
directly in this paper, do not have to be created by brane collisions. They can exist in these vacua 
independently of such considerations.

The outline of this paper is as follows. In section II we shall introduce the higher dimensional 
action upon which our analysis is based. We will review Stromingers solution \cite{Strominger:et} 
describing a 
gauge five brane in flat space and shall then proceed to generalise this solution to give an 
approximate configuration upon which we can carry out our dimensional reduction. In section III we 
proceed to the calculation of the effective action in four dimensions, first outlining a subtlety 
associated with promoting the integration constants of the background configuration to be four 
dimensional moduli fields, and then performing the dimensional reduction necessary to obtain the 
four dimensional theory. In section IV we present our results, in particular couching our findings 
in terms of a K\"{a}hler potential. In section V we comment on possible directions of future work.

Our index conventions are as follows. Indices $\mu$, $\nu$ ... and $M$, $N$ ... label world volume 
directions of the gauge five brane with $\mu$, $\nu = 0,1,2,3$ 
and $M$, $N = 4,5$. The $\mu$ directions will eventually correspond to four dimensional 
uncompactified space while the $M$ directions will be associated with a holomorphic 2 cycle in a 
Calabi-Yau threefold.   Indices $A$, $B$ ...  label the transverse dimensions with $A$, $B = 6,7,8,9$
. These directions will eventually correspond to directions in the compactified space transverse to 
the gauge five brane. We shall use indices $a$, $b = 4,..,9$ to denote a general direction in 
the internal space.


\section{Higher Dimensional Action and Background Solution}
\label{BG}

 Our starting point is the low energy effective action of the $E8 \times E8$ heterotic string. 
This action also 
provides an effective description of ten dimensional heterotic M-theory \cite{Lukas:1998ew} and, with a 
suitable change of 
gauge group, the weakly coupled $SO(32)$ heterotic string at low energies. Thus the discussion and 
results presented
in this paper are equally valid in these corners of the M-theory moduli space. 
Working with this description
 of these theories is valid to the approximations we shall be making and results in
considerable simplification as compared to carrying out the analysis, for example, in the eleven 
dimensional picture of heterotic M-theory. One obvious simplification as compared to that case
 is that we no longer have to worry about warping 
in the Ho\v{r}ava-Witten orbifold direction as this is already included in the effective theory we 
are using to the approximation we require. The action is given by,

\begin{eqnarray}
\label{10Daction}
S_{10} = \frac{1}{2 \k^2_{10D}} \int d^{10}x \sqrt{-g} \;  e^{2\phi} \left( -R - 4(\partial \phi)^2 + 
\frac{1}{3} H^2 + \frac{\a'}{30} \textnormal{Tr} F^2  + ...\right)
\end{eqnarray}
where the field strength H takes the usual form,
\begin{eqnarray}
\label{10DHdef}
H= dB -\frac{\a'}{30} \omega_{3YM} + ...\; .
\end{eqnarray}
The $...$'s here express the fact that we have dropped some terms which we will not need, for the 
particular calculation we are interested in, to our approximations.
The traces in this expression are in the adjoint of $E_8 \times E_8$.The
Chern-Simons three form associated with the $E_8 \times E_8$ gauge fields is denoted by $\omega_{3YM}$
 and B is a two form 
potential. The field strength of the $E_8 \times E_8$ gauge fields is $F$ and $\phi$ is the 
ten dimensional dilaton. The action is valid to first order in $\a'$ which is the order we will be 
working to throughout this paper.

\subsection{Gauge five brane solution in asymptotically flat space.}
\label{strominger}

There is a solution of this theory due to Strominger \cite{Strominger:et} which describes a gauge five 
brane in 10 dimensional asymptotically flat space (the 11 dimensional counterpart of this solution in 
the heterotic M-theory case was given in \cite{Lalak:1997ti}). 
While as it stands this is obviously not a suitable 
background solution for our purposes it will be very important in the following analysis and so we 
take the time to describe  it in some detail. The solution assigns values to the ten dimensional 
fields as follows.

\begin{eqnarray}
\label{BGA}
A_B &=& \sigma_{\gamma} \theta^{\gamma} \frac{2 i \rho^2 r^C \bar{\sigma}_{CB}}{R^2(R^2+ \rho^2)} 
	 \bar{\sigma}_{\delta} \theta^{\delta} \\ 
e^{-2 \phi} &=& e^{-2 \phi_0} \left( 1 + 8 \a' \frac{R^2 + 2 \rho^2}{R^2 + \rho^2} \right) \\  
ds^2_{10} &=&  \eta_{\mu \nu} dx^{\mu} dx^{\nu} + \delta_{ MN }dx^M dx^N +  e^{2 \phi_0 - 2 \phi} 
(\delta_{AB} dx^A dx^B) \\ 
H_{ABC} &=&  \epsilon_{ABC}^{\;\;\;\;\;\;\;\;\;D} \partial_D \phi \; 
\end{eqnarray}
In these expressions $\phi_0$, $\rho$ and $\theta^{\gamma}$ are constants and 
we make use of the following definitions.

\begin{eqnarray}
\label{r}
(r^C) &=& (x^6, x^7 , x^8 , x^9) \;  \\
R^2 &=& \delta_{AB} r^A r^B \; 
\end{eqnarray}
We also have the constraint 

\begin{eqnarray}
\label{thetaconstraint0}
\sum_{\gamma =1}^4 (\theta^{\gamma})^2 = 1 
\end{eqnarray}
on the quantities $\theta^{\gamma}$.

Indices $\mu$, $\nu$ ... and $M$, $N$ ... label world volume directions with $\mu$, $\nu = 0,1,2,3$ 
and $M$, $N = 4,5$. Indices $A$, $B$ ...  label the transverse dimensions with $A$, $B = 6,7,8,9$. 
We have split the world volume indices into two groups like this and have introduced $r$ and $R$ to 
make our notation compatible with the discussion in later sections when we will wrap two of the 
worldvolume directions of the gauge five brane up on a holomorphic cycle in a Calabi -Yau threefold. 
We define $\sigma_A = (1_{[2] \times [2]} , i \vec{\tau})$ where $\tau^i$, 
$i = 1,2,3$ are the Pauli matrices. The Hermitean conjugate matrices are 
$\bar{\sigma}_A = (1_{[2] \times [2]} , -i \vec{\tau})$ and we define the self dual and anti self dual
 two index objects, 
$\sigma_{AB} = \frac{1}{4} (\sigma_A \bar{\sigma}_B - \sigma_B \bar{\sigma}_A)$ and 
$\bar{\sigma}_{AB} = \frac{1}{4} (\bar{\sigma}_A \sigma_B - \bar{\sigma}_B \sigma_A)$. The 
completely antisymmetric symbol in 4 dimensions is denoted $\epsilon^s_{ABCD}$, and 
$\epsilon^s_{6789} = 1$. The associated tensor is denoted $\epsilon_{ABCD}$.

It should be noted that while this solution is in a singular gauge \cite{Dorey:2002ik} (the gauge 
field given above is 
divergent at $r^C =0$) any physical (i.e. gauge invariant) quantity associated with it is everywhere 
finite.

The solution is accurate, as is the action we have presented, to first order in $\alpha'$. 
It describes a soliton with 6 worldvolume dimensions and 4 transverse ones. The object is, 
at its core, a Yang Mills instanton embedded within some $SU(2)$ subgroup of the $E_8$ associated 
with the fixed plane on which it exists in the higher dimensional picture. This Yang Mills 
configuration has, as is well known, several collective coordinates. These are integration constants 
of the solution which describe flat directions in the objects moduli space. For example the object 
has some finite width in the four transverse directions which is determined by the constant $\rho$ 
in the solutions given above. The instanton has an orientation within the $SU(2)$ which is 
determined by the parameters $\theta^{\gamma}$. Although there are four $\theta$'s they are subject 
to one constraint (\ref{thetaconstraint}), and so the size and $SU(2)$ orientation together makes a 
total of four 
parameters which span the so called centred moduli space of the instanton. In addition to the 
centred moduli space the object has 4 collective coordinates which describe its motion in the 
transverse space and 112 zero modes associated with the embedding of $SU(2)$ within $E_8$. We shall 
concentrate on the centred moduli space in this paper and leave the 
analysis of these extra degrees of freedom for future work. It is fairly easy to see why this is a 
consistent thing to do, at least in the case of the translation moduli. The first non-trivial test 
that it is possible to consistently truncate off 
the position moduli is given by the observation \cite{Dorey:2002ik} that the moduli space of a pure 
Yang Mills instanton 
factors into a product of the centred moduli space and the space of the position moduli. Since this 
result must be regained in the limit where we 'freeze' the other moduli this is a necessary 
condition for our truncation to be consistent. The real test for consistent truncation however comes 
from checking that the position moduli enter the effective four dimensional theory bi-linearly. 
We have indeed checked that this is the case and so we are justified in truncating to the centred 
moduli space.

In any case one may wonder whether these gauge bundle moduli span the moduli space of the gauge 
five-brane or whether there are other moduli associated with the 'NS' dressing. As was demonstrated 
in \cite{Strominger:et} however, given the gauge field configuration presented in (\ref{BGA}) we 
can determine 
the 
gravitational dressing, also given in equations (\ref{BGA}), and no more integration constants 
appear 
(although different discrete choices are possible \cite{Callan:dj,Callan:1991ky,Duff:1994an}). 
The fact that this 
object is based upon a self-dual solution to the Yang-Mills equations will be of central importance 
when we come to talk of generalising the work presented in this paper. Their exists a powerful tool 
for obtaining such configurations, in the form of the ADHM construction \cite{Dorey:2002ik}, 
which we can use as a 
starting point for the analysis of more complicated situations.

\subsection{Wrapping up the gauge five brane}

As we have already mentioned, the solution given in the previous section is, as it stands, of no use 
for our 
current purposes as the gauge five-brane described therein lives in asymptotically flat ten 
dimensional space. We wish to describe a situation where the manifold we are working on is not 
${\cal M}^{10}$ but ${\cal M}^4 \times X$ where X is a compact six dimensional manifold. To preserve 
${\cal N}=1$ supersymmetry in four dimensions X has to have $SU(3)$ holonomy with respect to the 
generalised connection including the three form field strength \cite{Strominger:uh}. In addition, 
for ${\cal N}=1$
 supersymmetry and four dimensional Poincar\'{e} invariance we require that four of the world volume 
directions of our gauge five brane span ${\cal M}^4$ with the remaining two wrapping a holomorphic 
two cycle within $X$ \cite{Witten:1996mz}.

Now obtaining an exact solution describing a compact Calabi-Yau manifold with an associated gauge 
bundle which includes a piece of the background configuration which can be identified as a wrapped 
gauge five brane is, as was mentioned in the introduction, beyond the capabilities of current 
technology. However we can obtain an ${\it approximate}$ solution which describes such a situation 
and which is valid only near the cycle which the five-brane wraps. We shall find in the next section 
that this approximate solution is all that we require, as might have been expected on physical 
grounds, to calculate the terms due to the presence of the five brane in the desired effective 
action.

So how do we construct this approximate solution? We shall really only require one property of our 
compactification manifold in order to construct such an approximation as a generalisation of 
the solution we have already encountered. That property is,

\begin{itemize}
\item Near the 2 cycle which the gauge five brane wraps the compact space may be written as 
$X= {\cal C}_2 
\times {\cal C}_4$, where $ {\cal C}_2$ is a Riemann surface and $ {\cal C}_4$ is a complex four 
dimensional space.
\end{itemize}

However, in order to demonstrate the existence of Calabi-Yau three folds with this property and to 
give us a concrete context in which to carry out our calculation we shall concentrate on the class 
of compact metric configurations which can be obtained by blowing up six dimensional orbifolds of 
$SU(3)$ holonomy. Note '$SU(3)$ holonomy' here denotes the holonomy of the metric neglecting the 
back reaction due to the presence of the gauge five brane. Our analysis, of course, includes this 
back reaction and so our full metric is not of $SU(3)$ holonomy.

We shall therefore start with some six dimensional orbifold which in addition to having SU(3) 
holonomy has the factorisation property mention above near the two cycle on which we shall wrap our 
gauge five brane. In order to simplify the following calculation we shall also require that the 
point group of our orbifold is such as to project out off-diagonal metric moduli. This is not 
necessary for our method to work but results in considerable simplification of the calculation, in 
particular in keeping the number of moduli we have to deal with at a manageable level, while still 
retaining the essential ingredients we are interested in. 

An example of an orbifold which has all three of these properties is a $Z_8 - I$ Coexeter orbifold 
with an $SO(5) \times SO(9)$ lattice \cite{Bailin:nk}. The orbifold is constructed as follows. We 
define three complex coordinates.

\begin{eqnarray}
z_1 &=& x^4 + i x^5 \\
z_2 &=& x^6 + i x^7 \\
z_3 &=& x^8 + i x^9
\end{eqnarray}
To construct the orbifold we start with flat space spanned by these complex coordinates. We then 
make identifications under the point group. In our case the action of the point group can be written 
as follows.

\begin{eqnarray}
z_1 &\rightarrow& e^{2 \pi i \frac{1}{4}} z_1 \\
z_2 &\rightarrow& e^{2 \pi i \frac{1}{8}} z_2 \\
z_3 &\rightarrow& e^{-2 \pi i \frac{3}{8}} z_3 
\end{eqnarray}

As we have said, this particular choice is an example where the off-diagonal metric 
moduli are projected out by the orbifolding. This can be seen by considering the action of the 
point group on an off-diagonal component of the metric such as $g_{z_1 \bar{z}_2}$.

Having gauged the point group we then perform identifications under a lattice to 
obtain a compact manifold. We use the root lattice of $SO(5) \times SO(9)$, which 
is compatible with our choice of point group and with 
our need for a suitable two cycle on which to wrap the brane \cite{Bailin:nk}. 
 ${\cal C}_2$ will be identified with the 
space which is 
compactified by modding out by the $SO(5)$ lattice and ${\cal C}_4$ with the space which is modded 
out by the $SO(9)$ lattice. The orbifold fixed loci in this compactification are then blown up using some 
appropriate resolutions \cite{Polchinski:rr} leaving us with a class of 
Calabi-Yau 
manifolds with the desired properties.

Wrapping the gauge five brane solution up on a 2 cycle is then simple. We shall work in the 
'downstairs' picture where we simply consider the fundamental region of the orbifold. We choose the 
size of the fixed loci blow ups and the gauge five brane width to be much smaller than the overall
 size of the orbifold. We then choose a holomorphic 2 cycle to wrap around which is determined by 
choosing a point in ${\cal C}_4$ which is far from any of the orbifolds resolved fixed loci and the 
rest of the bundle which is also assumed to be localised (perhaps in the form of more gauge five 
branes). The solution given in section \ref{strominger} can then be generalised to wrap this cycle 
with the trivial modification of making the identifications that turn two of the world 
volume directions into the cycle (the solution has symmetries which are compatible with this). 

All we then have to do is use coordinate transformations to introduce the constants which will form 
the metric moduli of $X$ and introduce constants which 
will become the imaginary parts in their complexifications. We then have a 
solution which is valid near to the two cycle which will be appropriate to use in our dimensional 
reduction. Of course the solution differs substantially from the real situation we are interested in 
far 
from the gauge five brane in the transverse space, our approximate solution being asymptotically 
flat in these directions. However as we shall see in the next section we can show that this 
approximate solution, valid in this restricted volume to an approximation to be made more concrete 
later, is all we require to compute the effective theory we desire. 

The coordinate transformations we use to introduce the metric moduli are, in real coordinates,

\begin{eqnarray}
x^A &\rightarrow& \cV_{(A)}^{\frac{1}{6}} x^A \;
\end{eqnarray}

The constant two form potential contributions which will form the completion of the complex volume 
moduli can simply be added to the configuration and it will remain a solution. These can be seen in 
the equation for the two form below (they are the $\chi$'s). It should be noted 
however that the off-diagonal components of the two form are projected out for our choice of point 
group in exactly the same way as we saw above for the metric moduli. 

Combining all of this information we may now write down our approximate background solution including 
all of the necessary integration constants.

\begin{eqnarray}
\label{BGgaugefield0}
A_B &=& \sigma_{\gamma} \theta^{\gamma} \frac{2 i \rho^2 r^C \bar{\sigma}_{CB}}{R^2(R^2+ \rho^2)}  
\bar{\sigma}_{\delta} \theta^{\delta} \\
e^{-2 \phi} &=& e^{-2 \phi_0} \left( 1 + 8 \a' \frac{R^2 + 2 \rho^2}{R^2 + \rho^2} \right) \\
ds^2_{10} &=& f g_{\mu \nu} dx^{\mu} dx^{\nu} + \cV_3^{\frac{1}{3}} \delta_{ MN }dx^M dx^N +  e^{2 
\phi_0 - 2 \phi} (\cV_{(A)}^{\frac{1}{3}} \delta_{AB} dx^A dx^B) \\
B_{AB} &=& B^{bg}_{AB} + \frac{1}{6} \chi_{(1)} \Pi^{(1)}_{AB} + \frac{1}{6} \chi_{(2)} 
\Pi^{(2)}_{AB} \\
B_{MN} &=& B^{bg} + \frac{1}{6} \chi_{(3)} \Pi^{(3)}_{MN}  
\end{eqnarray}
where $\rho$ has been rescaled in the same way as the coordinates and we have introduced the notation
 $\cV_{(A)}$. This is taken to mean $\cV_{(1)} = \cV_{(2)} = \cV_1$, $\cV_{(3)} = \cV_{(4)} = \cV_2$.
 We have chosen to use this notation as each volume modulus, while associated with one component of 
the metric in complex coordinates ($g_{z_i \bar{z}_i}$ for $i=1..3$), is associated with two real 
coordinates. 
We have denoted three 
harmonic two forms on a flat orbifold of our type as (in the absence of the instanton)
 $\Pi^{(1)}_{ab} = (+1|_{a=1,b=2},-1|_{a=2,b=1})$,
 $\Pi^{(2)}_{ab} = (+1|_{a=3,b=4},-1|_{a=4,b=3})$ and  
$\Pi^{(3)}_{ab} = (+1|_{a=5,b=6},-1|_{a=6,b=5})$.

Finally $B^{bg}$ is the order $\alpha'$ background contribution to the two form due to the presence 
of the five brane. 

We now have the following definitions for $r^C$ and $R$.

\begin{eqnarray}
\label{r2}
(r^C) &=& (\cV_1^{\frac{1}{6}} x^6, \cV_1^{\frac{1}{6}} x^7 , \cV_2^{\frac{1}{6}} x^8 , 
\cV_2^{\frac{1}{6}} x^9) \; , \\ 
R^2 &=& \delta_{AB} r^A r^B \; 
\end{eqnarray}
We also still have the constraint on the $\theta^{\gamma}$'s.

\begin{eqnarray}
\label{thetaconstraint}
\sum_{\gamma =1}^4 (\theta^{\gamma})^2 = 1 \; .
\end{eqnarray}

$\cV_3$ is a metric modulus associated with the size of the 2 cycle the five brane wraps, i.e it is 
the size modulus associated with ${\cal C}_2$, and $\chi_3$ is its corresponding axion. $\cV_1$ and 
$\cV_2$ are metric moduli associated with the size and shape of the four dimensional transverse space
 ${\cal C}_4$ and $\chi_1$ and $\chi_2$ are their corresponding axions. $f$ is a Weyl rescaling 
factor which will be chosen by demanding that the Einstein Hilbert term in four dimensions is 
canonically normalised in terms of the four dimensional metric $g_{\mu \nu}$.

Now we have this approximate solution one might naively think that it is easy to compute the four 
dimensional effective action describing the gauge five brane. The 'usual' procedure would be to 
take integration constants in this approximate solution and promote them to be four dimensional 
fields. One would then take the resulting configuration, substitute it into the higher dimensional 
action and integrate out the six compactified dimensions. If our physical argument that the 
effective theory can be calculated without knowing what happens far from the gauge five brane's world 
volume is true then we should be able to do all of this without picking up non-negligible 
contributions from the part of the transverse space on which we cannot trust our solution. We would
 then end 
up with a four dimensional effective action which would be valid to some well controlled 
approximations. 
In fact we shall see that, while this is broadly speaking how the calculation proceeds, a few 
subtleties arise which we have to deal with before we can obtain our result.


\section{The Four Dimensional Moduli Space Effective Action}
\label{modspace}

\subsection{Promoting constants in the background solution and the inclusion of compensators}
\label{promotion}

	The first thing we have to do in the calculation of the effective action is to take the 
constants which will become the moduli we wish to describe and replace them with four dimensional 
fields. We immediately encounter a subtlety in doing this, albeit one that is well known in other 
contexts 
\cite{Dorey:2002ik}. Consider the following contribution to the gauge field.

\begin{eqnarray}
\label{compensator}
A_{\mu} = \Omega^{(A)}_{{\it m}} \partial_{\mu} {\it m}
\end{eqnarray}

Here ${\it m}$ is some modulus, a sum over moduli being implied on the right hand side, and we 
recall that ${\mu}$ labels an external four dimensional direction. 

Now let us construct a configuration by taking some background solution, promoting its integration 
constants to be four dimensional fields, and adding to it contributions to the gauge field 
of the form given in equation (\ref{compensator}). When we then take the four dimensional fields, {\it m},
 to be constant we recover the background solution for any value of the so called compensators 
$\Omega^{(A)}_{{\it m}}$. 
Similar points could be made for the metric  and 2 form fields. We could have,

\begin{eqnarray}
g^{comp}_{\mu a} = \Omega^{(g)}_{(a |{\it m}|} \partial_{\mu)} {\it m} \\
B^{comp}_{\mu a} = \Omega^{(B)}_{[a |{\it m}|} \partial_{\mu]} {\it m} \; .
\end{eqnarray}
The reason for only considering one four dimensional index in these expressions will become clear in 
a moment. It is clear then that we need some way of determining what values 
these compensators should take when we promote the integration constants of our background solution to 
be four dimensional moduli fields.

The idea of a moduli space approximation such as the one we are going to make is that a solution to 
the resulting four dimensional effective action can be raised, using the ansatz used for dimensional 
reduction, to give a solution to the higher dimensional equations of motion to some approximations. 
One
 of the approximations that is always made is the slowly changing moduli approximation. This is that 
the higher dimensional equations of motion will only be solved by such a configuration up to second 
order in four dimensional derivatives. Expanding the higher dimensional equations of motion in powers
 of four dimensional derivatives we obtain the following.

\begin{itemize}
\item {\bf Zeroth Order}: These equations are simply the background equations of motion - if we have 
chosen our background configuration correctly these are automatically satisfied to our 
approximations.

\item {\bf First Order}: These are the equations which determine the compensators - see below.

\item {\bf Second Order}: These are the higher dimensional manifestations of the moduli equations of 
motion.

\end{itemize}

Thus the first order equations determine the compensators. This is best demonstrated by an example 
so let us consider the gauge field compensators. The contribution to the  higher dimensional 
equation of motion for the gauge field at first order in four dimensional derivatives is,

\begin{eqnarray}
\label{precompeqn}
\cV_{(a)}^{-\frac{1}{3}} \cV_{(b)}^{-\frac{1}{3}} \frac{1}{3} \partial_{\mu} \chi_{(a)} 
\Pi^{(a)}_{ab} 
F_{ab} \alpha' + \alpha' {\cal D}_a \left( \cV_{(a)}^{-\frac{1}{3}} \left[ - \partial_{\it m}  
A_{a} \partial_{\mu} {\it m} + {\cal D} \Omega^{(A)}_{{\it m}} \partial_{\mu} {\it m} \right] 
\right) = 0 \; .
\end{eqnarray}

Here ${\cal D}$ is a gauge covariant derivative and summation is implied over repeated indices. We 
recall that the 
indices $a,b...$ cover the entire Calabi-Yau threefold, i.e. 
$a,b = 4,..,9$.

Now for this equation to be satisfied the coefficient of $\partial_{\mu} {\it m}$ has to vanish for 
each ${\it m}$. This leads to a separate equation for each gauge compensator.
For ${\it m} \neq \chi$ we have,
\begin{eqnarray}
\label{normcomp}
{\cal D}_a \left(  \cV_{(a)}^{-\frac{1}{3}} {\cal D}_a \Omega^{(A)}_{{\it m}} \right) = {\cal D}_a 
\left(\cV_{(a)}^{-\frac{1}{3}}  \partial_{{\it m}} A_a \right) \;.
\end{eqnarray}
However, for ${\it m} = \chi_{(A)}$ we have,
\begin{eqnarray}
\label{chicomp}
{\cal D}_a \left(  \cV_{(a)}^{-\frac{1}{3}} {\cal D}_a \Omega^{(A)}_{{\chi_{(a)}}} \right) = - 
\cV_{(a)}^{-\frac{1}{3}}  \cV_{(b)}^{-\frac{1}{3}} \frac{1}{3}  \Pi^{(a)}_{ab} F_{ab} \;.
\end{eqnarray}

We see from the structure of the source terms in these equations that the general rule is that if 
the background gauge field depends on a certain modulus then we have to include a gauge field 
compensator for that degree of freedom when we promote integration constants to be moduli. The 
exceptions to this rule are the fields $\chi_{(i)}$ ($i=1..3$) 
which have an additional source term causing them 
to give rise to compensators even if (as is indeed the case) the background gauge field is not 
dependent on them. Given these equations and our approximation to the background solution given in 
the previous section we can calculate expressions for the gauge field compensators. Of course, as 
before, our results will only be valid near the world volume of the gauge five brane. In addition we 
will need two boundary conditions to fix the compensators uniquely (due to them being determined by 
second order differential equations). The next question then is how do we choose these boundary 
conditions.

The first boundary condition is simply that when we compute any physical (i.e. gauge invariant) 
quantity we require it to be non-singular at the core of the gauge five-brane. This condition is 
enough to determine one of the integration constants that is present in the general solutions to 
equations (\ref{normcomp}) and (\ref{chicomp}). The second boundary condition is that we require 
that, in our 
approximate solution 
where the transverse space is asymptotically flat, the compensator does not diverge at large 
distances from the gauge five brane. This second condition is clearly necessary if our approximation 
is to work but can also be justified on physical grounds. The need for compensators here is a direct 
consequence of the presence of the five brane - they are not needed in the case where we ignore the 
instanton moduli. As they are sourced by the gauge five brane we would not expect the compensators  
to diverge as we go away from the soliton core in the transverse space.

Using our equations (\ref{normcomp}) and (\ref{chicomp}) and these boundary conditions we can then 
compute 
the approximations to the gauge field compensators associated with the configuration presented in 
the previous section. The result is given below for those moduli whose compensators do not obviously 
vanish.

\begin{eqnarray}
\label{compensators}
\Omega^{(A)}_{\rho} &=& 0 \\ 
\Omega^{(A)}_{\theta_{\gamma}} &=& \frac{-i \rho^2 \sigma_{\gamma} \bar{\sigma}_{\delta} 
\theta^{\delta}}{r^2 + \rho^2} \\
\Omega^{(A)}_{\cV_1}&=& \sigma_{\gamma} \theta^{\gamma} \frac{i}{3 \cV_1} \left(  \frac{ \left[ 
\bar{\sigma}_C r^C \sigma_{B7} r^B r^7 \sigma_D r^D  + \bar{\sigma}_C r^C \sigma_{B6} r^B r^6 
\sigma_D r^D \right]}{ R^2 ( R^2 + \rho^2)}  \right. \\ \nonumber  &&\left. - \frac{\left[ \bar{\sigma}_6 
r^6 
\sigma_C r^C + \bar{\sigma}_7 r^7 \sigma_C r^C\right]}{2 R^2} + \frac{((r^6)^2 + (r^7)^2)}{2 R^2} 
\right) \bar{\sigma}_{\delta} \theta^{\delta} 
\end{eqnarray}
\begin{eqnarray} 
\Omega^{(A)}_{\cV_2}&=& \sigma_{\gamma} \theta^{\gamma} \frac{i}{3 \cV_2} \left(  \frac{ \left[ 
\bar{\sigma}_C r^C \sigma_{B9} r^B r^9 \sigma_D r^D  + \bar{\sigma}_C r^C \sigma_{B8} r^B r^8 
\sigma_D r^D \right]}{ R^2 ( R^2 + \rho^2)} \right. \\ \nonumber 
&&\left.  - \frac{\left[ \bar{\sigma}_3 r^3 
\sigma_C r^C + \bar{\sigma}_9 r^9 \sigma_C r^C\right]}{2 R^2} + \frac{((r^8)^2 + (r^9)^2)}{2 R^2} 
\right) \bar{\sigma}_{\delta} \theta^{\delta}\\ 
\Omega^{(A)}_{\chi_1}&=& \sigma_{\gamma} \theta^{\gamma} \frac{i \rho^2}{3} \frac{ \bar{\sigma}_C r^C 
\bar{\sigma}_{67} \sigma_D r^D}{\cV_1^{\frac{1}{3}} R^2 (R^2 + \rho^2)}  \bar{\sigma}_{\delta} 
\theta^{\delta}\\ 
\Omega^{(A)}_{\chi_2}&=&\sigma_{\gamma} \theta^{\gamma} \frac{i \rho^2}{3} \frac{ \bar{\sigma}_C r^C 
\bar{\sigma}_{89} \sigma_D r^D}{\cV_2^{\frac{1}{3}} R^2 (R^2 + \rho^2)} \bar{\sigma}_{\delta} 
\theta^{\delta}
\end{eqnarray}

Thus when we promote integration constants to moduli fields in our background solution, prior to 
dimensional reduction, we must include these compensators when we write down the gauge field using 
the form shown in (\ref{compensator}). It should be noted that, while our boundary conditions are 
enough to give 
isolated values to the integration constants, in one case at least a different discrete choice 
for the compensator is possible. The expression $\Omega^{(A)}_{\theta^{\gamma}} = -i \sigma_{\gamma} 
\bar{\sigma}_{\delta} \theta^{\delta}$ also fits all of the above criteria for our compensators. 
However, this choice turns the $\theta$'s into the parameters of a local gauge transformation and so 
they drop out of the four dimensional effective action altogether. In short this choice means we are 
not including enough true integration constants in our ansatz - the ansatz is incomplete.

We should say a few words about gauge invariance at this point. Despite the compensators being gauge 
dependent quantities it is easy to show that the result they contribute to, i.e. the moduli space 
metric, 
is a gauge invariant quantity. This is essentially because the compensators are parts of gauge fields
 and so transform appropriately. The action of course is made out of gauge invariant quantities, a 
point we will return to briefly later, and thus leads to an invariant moduli space metric for our 
result. 

We mentioned that we will also obtain compensators for the metric and 2 form field in very similar 
ways. Fortunately, for the calculation we are interested in in this paper, the explicit forms of 
these compensators are not required and so we shall not give them here.

Collecting all the information we have presented in this section together we can now write down an 
approximation to the promoted background solution, where all of the integration constants have been 
replaced with four dimensional fields, which is valid near the gauge five brane's world volume. This 
is
 the configuration which it is appropriate to use in performing a reliable dimensional 
reduction to obtain the effective theory that we desire.

\begin{eqnarray}
\label{BGgaugefield}
A_B &=& \sigma_{\gamma} \theta^{\gamma} \frac{2 i \rho^2 r^C \bar{\sigma}_{CB}}{R^2(R^2+ \rho^2)} 
\bar{\sigma}_{\delta} \theta^{\delta} \\ \nonumber 
A_{\mu} &=& \Omega^{(A)}_{{\it m}} \partial_{\mu} {\it m} \\
e^{-2 \phi} &=& e^{-2 \phi_0} \left( 1 + 8 \a' \frac{R^2 + 2 \rho^2}{R^2 + \rho^2} \right) \\
ds^2_{10} &=& f g_{\mu \nu} dx^{\mu} dx^{\nu} + \cV_3^{\frac{1}{3}} \delta_{ MN }dx^M dx^N +  
e^{2 \phi_0 - 2 \phi} (\cV_{(A)}^{\frac{1}{3}} \delta_{AB} dx^A dx^B) \\ \nonumber
&& + 2 \Omega^{(g)}_{(a |{\it m}|} \partial_{\mu)} {\it m} dx^a dx^{\mu}  \\
B_{AB} &=& B^{bg}_{AB} + \frac{1}{6} \chi_{(A)} \Pi^{(A)}_{AB} \\
B_{MN} &=& B^{bg} + \frac{1}{6} \chi_{(3)} \Pi^{(3)}_{MN}  \\
B_{a \mu} &=& \Omega^{(B)}_{[a |{\it m}|} \partial_{\mu ]} {\it m} \\ \label{bggaug2}
B_{\mu \nu} &=& \textnormal{constant}
\end{eqnarray}

\subsection{Calculating the moduli space effective action}
\label{calc}

We are now in a position to calculate the centred moduli space metric of the gauge 
five brane. Our procedure for this calculation comes in several parts.

Firstly we must make sure that we use all of the information we have at our disposal. By a careful 
examination of what we know about the nature of the exact background configuration being considered 
we can show that some of the terms that could contribute to the moduli space metric in fact do not.
 Although we 
do not have a complete solution to first order in $\alpha'$ describing the Calabi-Yau and its 
associated gauge bundle we do know a small amount of information about this 'complete solution' (as 
opposed to the approximate one which we do have which is valid near to the five brane's world volume).

\begin{itemize}
\item $X$ is compact.

\item Except for near to the resolved orbifold fixed points we know the solution for the NS fields 
everywhere on the 
compact manifold to zeroth order in 
$\alpha'$. It is simply the relevant 
zeroth order parts of the approximate ansatz given in equations (\ref{BGgaugefield})-(\ref{bggaug2}).

\item We know what order in $\alpha'$ various contributions come in at. This information can be 
gleaned from an examination of the equations of motion. 
For example gauge field 
compensators come in at zeroth order in $\alpha'$, as can be seen from equations (\ref{normcomp}) and 
(\ref{chicomp})
(and we do not need any higher order corrections to  
them to our approximations). The metric compensators, however, are first order, their zeroth order 
contribution vanishing. 

\item We have a small amount of information about the index structure of the full solution. 
In particular we know that 
the compensators that we have to include, to our approximations, have one four dimensional and one 
Calabi-Yau index.
\end{itemize}

We can use this information to eliminate terms (i.e. show that they are zero) in our effective 
action calculation as follows. 
We start with the ten dimensional effective action which we repeat here for convenience.

\begin{eqnarray}
\label{10Daction2}
S_{10} = \frac{1}{2 \k^2_{10D}} \int d^{10}x \sqrt{-g} \;  e^{2\phi} \left( -R - 4(\partial \phi)^2 + 
\frac{1}{3} H^2 + 2 \a' \textnormal{tr} F^2  +... \right)
\end{eqnarray}
We have changed our convention here so that all traces from now on will be taken in the fundamental 
of $SU(2)$. This results in the changed coefficient of the Yang-Mills kinetic term, for example, 
as compared with 
eq. (\ref{10Daction}).

We are now going to imagine that we have  a full solution which describes the Calabi-Yau 
compactification and the full gauge bundle, including the gauge five brane. We imagine plugging this 
solution into the different terms of the ten dimensional action and performing the integration over 
the compact space. Using only the information given above about this solution, and working only up 
to first order in $\a'$, we start to eliminate terms.

Let us start with the 10 dimensional dilatonic Einstein Hilbert term. Consider the $O(\a')$ 
contributions to this term due to the presence of the $O(\a')$ gravitational compensators. Using the 
known zeroth order parts of the reduction ansatz and the fact that  the gravitational compensators 
are of order $\a'$ we find that this term has the following form.

\begin{eqnarray}
\label{Rcontribgcomp}
&-&\frac{1}{2 \k ^2} \int d^{10}x \sqrt{-g} \;  e^{2 \phi} R |_{\textnormal{gravitational 
compensators}} 
\\ \nonumber &=&-\frac{1}{2 \k ^2} \int d^{10}x f^2 \cV_1^{\frac{1}{3}} \cV_2^{\frac{1}{3}} 
\cV_3^{\frac{1}{3}} e^{2 \phi_0}  \left( R_{IJ}|_{\a'=0} g_{comp}^{IJ} + \nabla^I\left( 
\nabla^J g^{comp}_{IJ} - g^{KL}_{\a'=o} \nabla_I g^{comp}_{KL} \right)\right)
\end{eqnarray}

Now we use the index structure of the compensator and the zeroth order solution to eliminate the 
first and last terms on the right hand side.

\begin{eqnarray}
&-&\frac{1}{2 \k ^2} \int d^{10}x \sqrt{-g} \; e^{2 \phi} R |_{\textnormal{gravitational 
compensators}} 
\\ \nonumber &=&-\frac{1}{ \k ^2} \int d^{10}x f^2 \cV_1^{\frac{1}{3}} \cV_2^{\frac{1}{3}} 
\cV_3^{\frac{1}{3}} e^{2 \phi_0}  \left(  \nabla^I\left( \nabla^J g^{comp}_{IJ} \right)\right)
\end{eqnarray}
Using some standard identities, some information about our zeroth order solution, remembering that 
our internal manifold is compact and working to first order in $\a'$ we obtain, 
\begin{eqnarray}
&-&\frac{1}{2 \k ^2} \int d^{10}x \sqrt{-g} \;  e^{2 \phi} R |_{\textnormal{gravitational 
compensators}} 
\\ \nonumber &=&-\frac{1}{ \k ^2} \int d^{10}x \sqrt{-g} \; e^{2 \phi} (\partial_{\mu} \phi_0) 
\left( \nabla_a g_{comp}^{\mu a} \right) \; .
\end{eqnarray}
This can be written, using the same information, as the integral of a divergence over the 
compactification manifold and so vanishes.

Note in particular in the above that the gravitational compensators do not contribute to $\sqrt{-g}$ 
to first order in $\a'$. Combinations such as 
$g^{a \mu}_{comp} \partial_{\mu} \phi \, \partial_a \phi$ are likewise higher order in $\a'$. 
This means finally that, as promised in section \ref{promotion}, the 
gravitational 
compensators completely drop out of the calculation and need not be considered in what follows.

Next we consider the $O(\a')$ contribution which is due to the 2 form compensators. The relevant 
term in the ten dimensional action is the following one.

\begin{eqnarray}
\label{Bcontribbcomp0}
\frac{1}{2 \k^2_{10D}} \int d^{10}x \sqrt{-g} \; e^{2\phi} \left( \frac{1}{3} H^2  \right)
\end{eqnarray}

This has a contribution due to the $B$ field compensators which becomes, when we use the fact that 
the 2 form compensators are $O(\a')$, work to $O(\a')$, use our knowledge of the zeroth order 
solution, and use eqn (\ref{10DHdef}),

\begin{eqnarray}
\label{Bcontribbcomp1}
\frac{1}{2 \k^2_{10D}} \int d^{10}x \sqrt{-g} \; e^{2\phi} \left( \frac{1}{3} H^2  
\right)|_{\textnormal{compensators}} =  \frac{1}{2 \k^2_{10D}} \int d^{10}x \cV_{(a)}^{-\frac{2}{3}} 
\left( \frac{1}{3} \partial_{\mu} \chi_{(a)} \Pi^{(a)}_{ab} ( dB^{comp}_{ab \mu} ) 2 \right)    \; .
\end{eqnarray}
Then we use our knowledge of the compensators index structure to obtain,

\begin{eqnarray}
\label{Bcontribbcomp2}
\frac{1}{2 \k^2_{10D}} \int d^{10}x \sqrt{-g} \; e^{2\phi} \left( \frac{1}{3} H^2  
\right)|_{\textnormal{compensators}} =  \frac{1}{2 \k^2_{10D}} \int d^{10}x \cV_{(a)}^{-\frac{2}{3}} 
\left( \frac{1}{3} \partial_{\mu} \chi_{(a)} \Pi^{(a)}_{ab} ( 4 \partial_a B^{comp}_{b \mu} ) 2 
\right)    \; .
\end{eqnarray}

This is again the integral of a total divergence over a compact manifold and so vanishes. 
It should be noted that since we are working with a gauge dependent term in singular gauge there is 
a risk of obtaining non-zero boundary terms when performing such integration by parts. Fortunately 
in our case we find that they all vanish. We shall return to the subject of gauge 
invariance of our results shortly. At any rate, we see that, as was the case with the gravitational 
compensators, the 2 form compensators drop out of our calculation and as was mentioned earlier we do 
not need to know their precise form, even in the region near to the gauge five brane's world volume.

Having eliminated the terms which arise due to the presence of the metric and 2 form compensators we 
can now go on to the next stage of our procedure to obtain a reliable calculation 
(one that we can show only depends on the form
 of the reduction ansatz near the gauge five brane) of the four dimensional effective action. This 
next stage is to deal with some problematic 
terms which {\it do} seem to depend on the form of the reduction ansatz in the region where it is not 
a good approximation to the true configuration. 

Consider for example the dilaton.

\begin{eqnarray}
\label{fulldill}
\phi = \tilde{\phi}_0 + \alpha' \tilde{\Phi}
\end{eqnarray}

Here, $\tilde{\phi}_0$ is some constant - the zeroth order part of the solution and $\alpha' 
\tilde{\Phi}$ is a correction due to the presence of the gauge five brane which is some complicated 
function of moduli and Calabi-Yau coordinates. This expression should be compared to eqn 
(\ref{BGgaugefield}) which
 describes our approximation to this solution near to the gauge five brane's world volume. An example 
of the problematic terms mentioned above would be an $O(\a')$ correction term to the four dimensional
 effective action which is proportional to the integral over the transverse space of $\tilde{\Phi}$. 
If we try and perform this integral using our approximate solution we find that we get a divergent 
answer - a clear sign that this correction term receives significant contributions from portions of 
the compactification manifold which are far from the gauge five brane's world volume where our 
approximate solution is valid. This might naively be thought to be an indication that our method can 
not be made to work. However, the problem can be solved by making a sensible definition of the 
constant/modulus $\tilde{\phi}_0$. We denote the average of this correction over the compact space 
as follows.

\begin{eqnarray}
\label{phiredef}
\frac{\left(\int d^6 x \alpha' \tilde{\Phi} \right)}{{V_{CY_3}}} = \alpha' <\tilde{ \Phi} >
\end{eqnarray}
Here, $V_{CY_3}$ is the coordinate volume of the Calabi-Yau. The trick is to define our dilatonic 
modulus so as to absorb our ignorance of the situation far from the gauge five-brane's world volume 
into this already arbitrary constant. We perform the following manipulations,

\begin{eqnarray}
\label{sensibledefphi}
\phi &=& \tilde{\phi}_0 + \alpha' <\tilde{\Phi}>  + \left[ \alpha' \tilde{\Phi} - \alpha'
<\tilde{\Phi}> \right] \\ 
 &\equiv& \phi_0 + \left[ \alpha' \Phi \right] \; .
\end{eqnarray}

In other words we absorb the average of the correction over the internal space into the definition 
of the constant $\phi_0$. Since this constant takes an arbitrary value in the background solution 
this is something we are perfectly entitled to do. 
With the dilaton in this form some of the potentially problematic terms which we couldn't calculate 
in the four dimensional effective theory vanish. For example, the analogue of the term we mentioned 
above is proportional to the integral over the transverse space of $\Phi$ which is zero.
 We will deal with the remaining troublesome terms with a similar trick involving the axions in a 
short while. But first let us make a digression to work out the four dimensional Einstein Hilbert 
term and so fix $f$, the Weyl rescaling factor in the metric.

Using all we have learned so far we find for the four dimensional Einstein Hilbert term,

\begin{eqnarray}
\label{4dEH}
- \frac{1}{2 \k_4^2} \int d^4 x \sqrt{ - g_4} f \cV_1^{\frac{1}{3}} \cV_2^{\frac{1}{3}} 
\cV_3^{\frac{1}{3}} e^{2 \phi_0} R_{4D} \; .
\end{eqnarray}
From this we see that for a canonically normalised Einstein Hilbert term in four dimensions we must 
choose,

\begin{eqnarray}
\label{fdef}
f = \cV_1^{-\frac{1}{3}} \cV_2^{-\frac{1}{3}} \cV_3^{-\frac{1}{3}} e^{-2 \phi_0} \; .
\end{eqnarray}

We now return to the examination of possible problematic terms in our calculation. The only
remaining terms which require knowledge of the field configuration far from the gauge five brane's 
world volume come from the $H^2$ term in the ten dimensional action. These terms are due to the 
background $B$ field configuration as opposed to the $B$ field compensators whose contributions to 
the effective
 action have already been shown to vanish. The terms are again of order 
$\a'$ and are given by,

\begin{eqnarray}
\frac{1}{2 \k^2} \int d^{10}x \cV_{(a)}^{-\frac{2}{3}} \frac{4}{3} \partial{\mu} \chi_{(a)} 
\Pi^{(a)}_{ab} \partial_{\mu} B^{\textnormal{bg}}_{ab}  \; .
\end{eqnarray}

This is an $O(\a')$ term as the background 2 form field is of that order. As with the terms we 
examined previously if we try and calculate this integral using our approximate solution we obtain 
an answer which depends on the form of the solution in the region of the transverse space where our 
approximation breaks down. As before we can deal with this problem by making a judicious, and 
perfectly legitimate, redefinition of our moduli. This time we redefine the geometrical axions to 
absorb the average of the background 2 form over the transverse space.

\begin{eqnarray}
\label{chiredef}
\Pi^{(c) ab} B_{ab} &=& \frac{1}{6} \chi_{(c)} 2 + \Pi^{(c) ab} B^{\textnormal{bg}}_{ab} \\ 
\label{chiredef2}
\Pi^{(c) ab} B_{ab} &=& \frac{1}{6} \left( \chi_{(c)} 2 +6 <\Pi^{(c) ab} B^{\textnormal{bg}}_{ab}>
\right) +  \left[ \Pi^{(c) ab} B^{\textnormal{bg}}_{ab} -  <\Pi^{(c) ab} B^{\textnormal{bg}}_{ab}> 
\right]
\end{eqnarray}

We redefine $\chi_{(c)}$ to be the terms in the first set of (non-square) brackets in the second line
 of equation (\ref{chiredef2}). As was the case in our redefinition of the dilatonic modulus this 
completely eliminates the problematic $O(\a')$ corrections by absorbing our ignorance of the 
'asymptotic' configuration into the arbitrary constant associated with the modulus.

We may now, finally, proceed to plug our approximate reduction ansatz into the remaining terms in the
 ten dimensional effective action. We then find that in the calculation of the 4d effective action - 
i.e. in performing the integration over the compact dimensions of the Calabi-Yau three fold - that we
 do not need to know the reduction ansatz to an accuracy beyond that provided by our approximation. 
In other words the terms involving five brane moduli only depend on the form of the reduction ansatz 
close to the gauge five brane's world volume.

Let us be a little more precise about what we mean by this. Consider patching our instanton into the 
transverse space by multiplying all of the fields by some smoothing function which is 1 inside some 
radius $r_1$ which includes the core, 0 outside some radius $r_2 > r_1$ and which smoothly 
interpolates between these values between these two radii. We find that our results would be 
unchanged by such a procedure - whatever the specific form of the smoothing function - provided that
 $r_1 >> \rho$. Our approximation neglects terms which are suppressed with respect to the ones which 
we keep by factors of $\frac{\rho^2}{r_1^2}$.

We find that all of the terms in the four dimensional action involving instanton moduli come from 
just two terms in the higher dimensional action. They are,

\begin{eqnarray}
\frac{1}{2 \k^2} \int d^{10} x \sqrt{-g} \; e^{2 \phi} \left( - \frac{4}{3} \a' \textnormal{d} 
B \omega_{3 YM} + 2 
\a' \textnormal{tr} F^2 \right)\; .
\end{eqnarray}

So all we have to do is to plug the zeroth order ansatz into this expression and perform the 
appropriate integrals.

\vspace{0.5cm}

The reader may be wondering what has happened to gauge invariance in all this. The 
$\textnormal{tr} F^2$ term is 
clearly gauge invariant. However at first site the $\textnormal{d} B \omega_{3 YM}$ term is not as 
it usually pairs 
with a $(\textnormal{d} B)^2$ term to form a gauge invariant object to first order in $\a'$. 
In fact we can show 
that, to the degree that is needed for our calculation, this term in the action is indeed gauge 
invariant in its own right to first order in $\a'$, up to our approximations. The proof is as follows.
Schematically the 2 form and gauge field change in the following manner under an  infinitesimal 
gauge transformation.

\begin{eqnarray}
\label{gaugetrans}
\delta A &=& d \Lambda + [A, \Lambda] \\
\Rightarrow  \delta \omega_{3YM} &=&  \textnormal{d}  \textnormal{tr} (\Lambda \textnormal{d} A)
  \\
\delta B &=& 2 \a' \textnormal{tr}(\Lambda \textnormal{d} A)
\end{eqnarray}
Here $\Lambda$ is an infinitesimal parameter describing the gauge transformation. This means that, to 
first order in $\a'$,
our $\textnormal{d} B \omega_{3YM}$ term undergoes a change under such a transformation proportional 
to,

\begin{eqnarray}
\label{gaug1}
\int d^6 x \; \a'  \textnormal{d} \left[ \textnormal{d}B 
\textnormal{tr} (\Lambda \textnormal{d} A) \right]\; .
\end{eqnarray}

This is the integral of a total derivative. If we are working at the level of our approximate 
solution this leads to two possibilities. Either the gauge transformation we are considering dies 
off sufficiently quickly away from the five brane's world volume that we can still perform our 
calculation reliably in the resultant gauge or it does not. In the first case due to the total 
derivative structure of the integrand in eq. (\ref{gaug1}) we find that our moduli space 
effective action 
indeed will not change if we perform this gauge transformation on our reduction ansatz. In the second 
case we will find that we simply cannot reliably perform the calculation in this gauge. Thus we find 
that in any gauge where we can actually compute the answer we desire our answer is unique and so our 
results are compatible with gauge invariance.

\section{The Result}
\label{result}

When we apply the procedure detailed in the previous section, taking into account all of the various
 subtleties, we find we can indeed reliably calculate the moduli space effective action of the gauge 
five brane without knowing the exact form of the solution far from the instanton core in the 
transverse space. The effective theory including the geometric moduli and the gauge five brane 
moduli is the following.

\begin{eqnarray}
\label{modaction}
S &=& \frac{1}{2 \k_4^2} \int d^4x \sqrt{-g} \left( -R + \frac{1}{2} (\partial \varphi)^2 + 
\frac{1}{2}\left( (\partial \beta_1)^2 + (\partial \beta_2)^2 +(\partial \beta_3)^2\right) 
+ e^{-2  \varphi} (\partial \sigma )^2  \right. \\ \nonumber && \;\;\;\;\;\;\;\;\;\;\; \left. 
+ \frac{2}{9} \left(e^{-2 \beta_1}( \partial \chi_1)^2 + e^{-2 \beta_2}( \partial \chi_2)^2 
+ e^{-2 \beta_3}( \partial \chi_3)^2 \right) \right. \\  \nonumber  && \;\;\;\;\;\;\;\;\;\;\; \left. 
 +  q_{G5}  \left[ \frac{}{} 8 e^{-\beta_1 -\beta_2} \left( ( \partial \rho)^2 + \rho^2 (\partial 
\theta)^2 \right)  + \rho^2 e^{-\beta_1- \beta_2} \left( ( \partial \beta_1 )^2 
+( \partial \beta_2)^2 \right)  \right. \right. \\ \nonumber &&\;\;\;\;\;\;\;\;\;\;\;  \left. 
\left.   -4 \rho e^{-\beta_1-\beta_2} \partial \rho \left( \partial \beta_1 + \partial \beta_2 
\right) -\frac{2}{9} \rho^2 e^{-\beta_1 - \beta_2}\left( e^{-2 \beta_1}( \partial \chi_1)^2 
+ e^{-2 \beta_2}( \partial \chi_2)^2 \right)  \right. \right. \\ \nonumber &&\;\;\;\;\;\;\;\;\;\;\;  
\left. \left.  + \frac{4}{9} \rho^2 e^{-2 \beta_1 -2 \beta_2} \partial \chi_1 \partial \chi_2
+\frac{8}{3} \rho^2 e^{-\beta_1- \beta_2} (\theta^2 \partial \theta^1 -\theta^1 \partial \theta^2 
+\theta^3 \partial \theta^4 - \theta^4 \partial \theta^3)(e^{- \beta_1} \partial \chi_1 
+ e^{- \beta_2 } \partial \chi_2)   \right] \right)
\end{eqnarray}

This expression employs the following definition.

\begin{eqnarray}
(\partial \theta)^2 = \sum_{\gamma=1}^4 (\partial \theta^{\gamma})^2
\end{eqnarray}
We have also introduced $q_{G5} = \frac{ \a' (2 \pi)^2}{V_{\textnormal{trans}}}$ and 
$V_{\textnormal{trans}}$ is the coordinate volume of the transverse space to the five brane. 
To get the result in this form, which exhibits the usual normalisations of low energy heterotic 
M-theory, we have made a number of field redefinitions. We have $\cV_{a} = e^{3 \beta_a}$, and 
$\phi_0 = \frac{1}{2} \varphi  - \frac{1}{2} ( \beta_1 + \beta_2 + \beta_3)$. In particular these 
choices ensure that the zeroth order (in $\a'$) kinetic terms have the usual form and normalisation.

The terms on the first line of the action then form the usual result for the four dimensional 
effective
 action of heterotic M-theory (or indeed weakly coupled heterotic string theory) 
accurate to first order in $\alpha'$. To recap  $\beta_3$ is a volume 
modulus for the 2 cycle 
our gauge five brane wraps and $\chi_3$ is its associated axion. $\beta_1$ and $\beta_2$ are volume 
moduli 
associated with the four dimensional transverse space and $\chi_1$ and $\chi_2$ are their associated 
axions. 
Finally $\varphi$ here is the four dimensional dilaton and $\sigma$ is its axion.

The remaining terms, which are contained within the square brackets, contain the kinetic terms of, 
and 
cross-
couplings to, the gauge five brane moduli. We recall that $\rho$ is the solitons width while the 
$\theta$'s 
describe its $SU(2)$ orientation. So, for example, if we want to describe how a gauge five brane 
spinning in 
$SU(2)$ space generates the axions $\chi_{(i)}$ from a four dimensional view point 
all we have to do is 
obtain an 
appropriate cosmological solution of this action.

 Let us recap the approximations we have made and thus when our effective action is valid. We have 
made all of the standard approximations employed in obtaining four dimensional actions in this 
context. These are the slowly changing moduli approximation, working to first order in $\a'$ (or more 
precisely to first order in $\e_w$ in the language of \cite{Lukas:1998hk} ) and ignoring towers of 
massive states (which corresponds to 'the other' $\e$ expansion of 
\cite{Lukas:1998hk}). The action also does not include contributions from non-perturbative 
corrections.

The new 
approximation that we have made here is that $\rho << V_{\textnormal{trans}}^{\frac{1}{6}}$ (there 
could be corrections to our action which are suppressed by powers of 
$\frac{\rho^2}{V_{\textnormal{trans}}^{\frac{1}{3}}}$). In addition, in order for our supergravity 
description to
 be valid we require $\rho^2 >> \a'$.

The first thing to notice about this result is that if we artificially 'turn off' gravity - i.e. if we 
drop all the metric 
moduli we obtain the lagrangian density $( \partial \rho)^2 + \rho^2 (\partial \theta)^2$. The moduli 
space 
associated with these kinetic terms is simply the usual centred moduli space of a single Yang Mills 
instanton \cite{Dorey:2002ik}. One might think, remembering the constraint (\ref{thetaconstraint}) on 
the 
$\theta$'s, that this moduli space is simply ${\cal R}^4$. However it is in fact 
$\frac{{\cal R}^4}{Z_2}$. 
This is because the physical situation is unchanged by the transformation $\theta^{\gamma} \rightarrow
 -\theta^{\gamma}$. Hence we should make the identification 
$\theta^{\gamma} = - \theta^{\gamma}$ which results in the moding out by $Z_2$ in the moduli space. 
Due to this identification the Yang Mills instanton moduli space has a conical singularity at the point
 where the size modulus vanishes. In fact it is clear that this conical singularity survives in our 
full 
result and this pathology at $\rho = 0$ is one of the indicators that we can not trust our effective 
action 
down to arbitrarily small gauge five brane widths. Another indication of this is that if we examine
 the curvature at the core of the object it diverges as we take the size to zero.
It should be noted that the moduli space of a single Yang Mills instanton is, from the above 
discussion, obviously 
hyper-K\"{a}hler as it is locally simply four dimensional flat space. 

We can see that the Yang Mills instanton moduli space 
is embedded within our result 
in a highly non-trivial way involving many prefactors and cross-terms. A very
 good check that we have got the calculation of all of these terms correct is that the result is 
compatible with ${\cal N} = 1$ supersymmetry. In other words we should check that the full 
moduli space is K\"{a}hler.

We have checked this in two different ways. Firstly we have directly calculated the holonomy of the
 moduli space. This is achieved by using the Riemann tensor to obtain the generators of the 
holonomy group. One may then count the number of linearly independent generators to obtain 
the dimensionality of the group. By examining the dimensionality of all the different possible 
subgroups of $SO(N)$, where $N$ is the dimensionality of the manifold under consideration,
 one can then in many cases show that only one subgroup has the dimensionality that we 
find - giving us the holonomy.

Now the moduli space is a direct product of a manifold spanned by $(\varphi,\sigma,\beta_3,\chi_3)$
 and one spanned by $\beta_1,\chi_1,\beta_2,\chi_2,\rho,\theta^{\gamma})$. We know from
 standard results that the first manifold is K\"{a}hler so we just need to find the holonomy of the 
second one. We find that its holonomy group fills out the entirety of $U(4)$ and hence our 
result is indeed compatible with ${\cal N} =1$ supersymmetry. We also see that the 
hyper-K\"{a}hler moduli space of the Yang-Mills instanton is reduced to being merely 
K\"{a}hler when embedded within this context.

The second way in which we can demonstrate that our result is K\"{a}hler is to write 
down a K\"{a}hler potential and complex structure which is associated with our component action 
(\ref{modaction}). We find the following.

\begin{eqnarray}
K &=& -\ln(S + \bar{S}) - \ln(T_1 + \bar{T}_1) - \ln(T_2 + \bar{T}_2) - \ln(T_3 + \bar{T}_3) 
+ \frac{16 \a' \left( |C_1|^2 + |C_2|^2 \right)}{\sqrt{\left( T_1 + \bar{T}_1 \right) 
\left( T_2 + \bar{T}_2 \right)}} \\
C_1 &=& e^{-\frac{\beta_1}{4} - \frac{\beta_2}{4}} \left( Y_1 + i Y_2 \right) \\
C_2 &=& e^{-\frac{\beta_1}{4} - \frac{\beta_2}{4}} \left( Y_3 + i Y_4 \right) \\
T_1 &=& e^{\beta_1} + \frac{2}{3} i \chi_1 + 4 \a' e^{\frac{\beta_1- \beta_2}{2}} 
\left( |C_1|^2 + |C_2|^2 \right) \\
T_2 &=& e^{\beta_2} + \frac{2}{3} i \chi_2 + 4 \a' e^{\frac{\beta_2- \beta_1}{2}}
 \left( |C_1|^2 + |C_2|^2 \right) \\
T_3 &=& e^{\beta_3} + \frac{2}{3} i \chi_3 \\ 
S &=& e^{\varphi} + \sqrt{2} i \sigma
\end{eqnarray}

Here we have defined the fields $Y_{\gamma}$ as $Y_{\gamma} = \rho \theta^{\gamma}$. 
We see that we have the usual K\"{a}hler potential of heterotic string/M-theory with an additional 
term, the final one, which is due to the presence of the gauge five brane. This additional term
 is just the K\"{a}hler potential for a simple Yang Mills instanton, modified by the addition
 of some factors of real parts of $T$ superfields. Similarly the definition of the $T$ 
superfields in terms of  component fields is just the usual one with a couple of modifications 
at $O(\a')$ due to the presence of the gauge five-brane. $C_1$ and $C_2$ are again 
just the usual Yang Mills instanton expressions modified by some overall factors of 
different $e^{\beta}$'s.

We can make a number of comments about the physics that follows from these results
 purely from an examination of the component action and K\"{a}hler structure.

\begin{itemize}
\item First of all it is clearly not consistent, in the case of the compactifications 
considered here, to take the universal case for the metric moduli in the presence of a 
generic changing instanton configuration. In other words, due to the non trivial factors of 
$e^{\beta_1}$ and $e^{\beta_2}$ in the gauge five brane moduli kinetic terms for example, it is 
inconsistent to set $\beta_1=\beta_2=\beta_3$ (however as we shall see shortly
 we can set $\beta_1=\beta_2$ if we make some compatible truncations of the other fields).

\item The dilaton and the size of the 2 cycle ($\varphi$ and $\beta_3$) do not feel the 
presence of the gauge five brane from the point of view of the four dimensional theory. 
The terms in which they appear are not modified from the results where the bundle 
moduli are ignored.

\item It can be seen from the last five terms in equation (\ref{modaction}) that it is not 
consistent to 
truncate off the gauge five brane moduli by setting them to be non-zero constants. In fact it 
is not possible to truncate them away by setting all the $Y$'s to
 zero either, even though they appear bilinearly in the above expressions. This is because setting
 all the $Y$'s to zero in this manner corresponds to setting $\rho$ to zero and as mentioned
 above our effective description is not valid in this region of moduli space. This result is in 
contrast to the case of matter fields for example \cite{Lukas:1998fg}. 
In that case there is no analogue of our $Z_2$ 
identification and so the matter fields (which appear bilinearly as well) can be truncated away by 
simply setting them to zero.

 Returning to the gauge five brane case, more complicated
 forms of truncation are possible in certain special cases. We shall see such a
 special case where we can clearly truncate off the instanton moduli, in certain combinations
 with other fields, in a moment. 
	
\item The form of the instanton corrections to the K\"{a}hler potential and complex structure is 
reminiscent of the analogous corrections obtained by the inclusion of matter fields. This is perhaps
 not particularly surprising given that there are some similarities in the origins of these four
 dimensional fields. 
It should be emphasised, however, that many qualitative differences exist in how these two types 
of moduli arise.
\end{itemize}

\vspace{0.5cm}

Now the result presented above appears fairly complicated and depends on a reasonably large number of 
fields (twelve real moduli). However, we can consistently truncate away many of the 
moduli to leave a 
simpler system with fewer degrees of freedom. For example the simplest non-trivial truncation which 
includes at least one instanton modulus is the following.

\begin{eqnarray}
\label{truncated}
K = -2 \ln\left( T + \bar{T} \right) + \frac{32 \a' |C|^2}{\left( T + \bar{T} \right)} \\ 
T = e^{\beta} + \frac{2}{3} i \chi + 8 \a' |C|^2 \\ 
C = e^{-\frac{\beta}{2}} \left( Y_1 + i Y_2 \right)
\end{eqnarray}
Here we have taken the situation where $\beta_1 = \beta_2 = \beta \;, \; \chi_1 = \chi_2 = \chi $, 
$\phi, \sigma, \beta_3$ and $\chi_3$ are taken constant, $Y_1 = Y_3$ and $Y_2 = Y_4$. 
We can obtain a 
component action from this K\"{a}hler potential and consistently truncate off the axions to obtain 
the 
following 4d theory.

\begin{eqnarray}
\label{simplecomponentS}
S = \frac{1}{2 \k^2} \int d^4 x \sqrt{- g} \left( -R + (\partial \beta)^2 + 8 q_5 e^{-\beta} (\partial
\hat{\rho})^2 \right)
\end{eqnarray}
We have defined $\hat{\rho}  = e^{-\frac{\beta}{2}} \rho$. Since this simplification was obtained 
via consistent truncation a solution to this simple theory corresponds to a solution to the full 
higher dimensional equations of motion to our approximations. One could obtain cosmological solutions 
to such an action with ease. These solutions are the simplest examples of what we need to make the 
brane collision scenarios mentioned in the introduction more complete. We have obtained such
 solutions 
and these will be presented as part of some future work \cite{jgal}.

\section{Further Work}
\label{further}

There are many ways in which this work can be used as a basis for future study. Here we will list a 
few of the more interesting possibilities.

\begin{itemize}
\item One could generalise the results presented here to the case where we consider more than one 
gauge five brane. It would not even be necessary to restrict  such a study to the case where the gauge
 five branes do not  overlap (a trivial modification of the above results). Such complicated 
situations are probably tractable due to the fact that we have a very powerful mechanism for 
obtaining self dual gauge field configurations in the form of the ADHM construction 
\cite{Dorey:2002ik}. This construction can provide us with analytic solutions for configurations
 containing many instantons. These can overlap, have different positions and sizes as well as 
different $SU(2)$ orientations. Following the procedure outlined in \cite{Strominger:et} we can
 then use these gauge field configurations as the core for a system of gauge five branes. The 
NS dressing can be determined once the gauge field configuration is known. Similar calculations
 to the one presented here could then be performed for these more complicated situations. 
Indeed by using a Kummer style construction for the Calabi-Yau threefold, such as the one we
 have employed here, and by taking the case where the gauge field background is entirely in 
the form of (either overlapping or not overlapping) gauge five branes one could write down a 
four dimensional theory which includes {\it all} of the moduli present in the compactification.

\item There are other parts of the gauge bundle which would yield to our approach. For example 
there is another object which takes a Yang-Mills instanton as its core - the so called symmetric
 solution \cite{Callan:dj,Callan:1991ky,Duff:1994an} (this is related to the discrete choices in
 determining the NS dressing that we mentioned earlier). We could equally well apply our method 
to this object and obtain the low energy effective theory which includes its moduli. Unlike our 
case this object is an example of a standardly embedded configuration. Another difference to the
 object we have considered in this paper is that the size modulus of the symmetric solution is 
quantised. This would mean that there would not be analogues of the particular continuous moduli
 we have considered here for that case.

\item One could try to obtain a more complete description of the gauge five brane's four dimensional 
effective theory by combining the kinetic terms described here with the work which has been done 
on obtaining non-perturbative potentials for gauge bundle moduli 
\cite{He:2003tj,Buchbinder:2002ji,Buchbinder:2002ic,Buchbinder:2002pr}. In particular it would be 
interesting to identify the moduli in these papers which correspond to those described here, for 
example the instanton size.

\item The action presented in this paper could be used to derive a number of different types of 
cosmological solutions. For example one could seek to describe the cosmological effects of a gauge 
five brane spreading out with time \cite{jgal} or spinning in $SU(2)$ space. Such solutions could 
be of critical importance in certain cosmological scenarios \cite{Khoury:2001wf,Bastero-Gil:2002hs}.

\item We have already described how our results could be used to improve the description of 
cosmological scenarios based upon small instanton transitions. 
However we would also like to stress that gauge five branes can live on the orbifold fixed planes
 irrespective of whether or not the system has undergone such phase transitions. Therefore it is 
possible to base scenarios purely on the dynamics of such objects. For example, if we were to 
include the position moduli of the five brane in our analysis we could imagine basing some kind
 of brane inflation scenario on gauge five branes and anti gauge five branes. This 
soliton-antisoliton inflation could potentially have some quite nice properties. For example
 when inflation ends with the collision of the instanton and anti instanton they would presumably
 annihilate in a manner which is describable within the regime of low energy field theory - 
both objects simply being made out of low energy fields \footnote{ Although some caution is called for 
with this statement given the results presented in \cite{Lima:1999dn}
for a situation which one would think would be subject to similar arguments. The two situations
 {\it are} different however. In particular in our case the two colliding objects would have no 
net five brane charge.}. The energy from such
 an annihilation could reheat the universe - the fact that the colliding objects are annihilating
 on an orbifold fixed plane presumably means that it would be natural for a sizable proportion of
 the resulting energy to be dumped into matter fields. 

In other words gauge five branes can be every bit as usefull in developing various scenarios as their
'fundamental' counterparts - and in addition these solitonic objects have extra attractive features 
such as variable widths and the fact that they are entirely built out of low energy fields.

\end{itemize}

In short it is now possible to start an analysis of the effect of certain types of gauge bundle moduli
 on different cosmological scenarios for the first time.

\section{Acknowledgments}

JG is supported by a Sir James Knott fellowship. AL is supported by a PPARC advanced fellowship.



\begin{thebibliography}{99}

\bibitem{Witten:1996mz}
E.~Witten,
``Strong Coupling Expansion Of Calabi-Yau Compactification,'' 
Nucl.\ Phys.\ B {\bf 471} (1996) 135
[hep-th/9602070].

\bibitem{Lukas:1998fg}
A.~Lukas, B.~A.~Ovrut and D.~Waldram,
``On the four-dimensional effective 
action of strongly coupled heterotic  string theory,''
Nucl.\ Phys.\ B {\bf 532} (1998) 43 
[hep-th/9710208].

\bibitem{Horava:1996qa}
P.~Ho\v{r}ava and E.~Witten,
``Heterotic and type I string dynamics from eleven dimensions,''
Nucl.\ Phys.\ B {\bf 460}, (1996) 506
[arXiv:hep-th/9510209].

\bibitem{Horava:1996ma}
P.~Ho\v{r}ava and E.~Witten,
``Eleven-Dimensional Supergravity on a Manifold with Boundary,''
Nucl.\ Phys.\ B {\bf 475}, (1996) 94
[arXiv:hep-th/9603142].

\bibitem{Green:mn}
M.~B.~Green, J.~H.~Schwarz and E.~Witten,
``Superstring Theory. Vol. 2: Loop Amplitudes, Anomalies And Phenomenology,''
{\it  Cambridge, Uk: Univ. Pr. ( 1987) 596 P. ( Cambridge Monographs On Mathematical Physics)}.

\bibitem{Banks:1996ss}
T.~Banks and M.~Dine,
``Couplings and Scales in Strongly Coupled Heterotic 
String Theory,''
Nucl.\ Phys.\ B {\bf 479} (1996) 173
[arXiv:hep-th/9605136].

\bibitem{Lukas:1999yy}
A.~Lukas, B.~A.~Ovrut, K.~S.~Stelle and D.~Waldram,
``The universe as a 
domain wall,''
Phys.\ Rev.\ D {\bf 59} (1999) 086001
[hep-th/9803235].

\bibitem{Brandle:2000qp}
M.~Brandle, A.~Lukas and B.~A.~Ovrut,
``Heterotic M-theory cosmology in four and five dimensions,''
Phys.\ Rev.\ D {\bf 63} (2001) 026003
[arXiv:hep-th/0003256].

\bibitem{He:2003tj}
Y.~H.~He, B.~A.~Ovrut and R.~Reinbacher,
``The moduli of reducible vector 
bundles,''
arXiv:hep-th/0306121.

\bibitem{Buchbinder:2002ji}
E.~Buchbinder, R.~Donagi and B.~A.~Ovrut,
``Vector bundle moduli and 
small instanton transitions,''
JHEP {\bf 0206} (2002) 054
[arXiv:hep-th/0202084].

\bibitem{Buchbinder:2002ic}
E.~I.~Buchbinder, R.~Donagi and B.~A.~Ovrut,
``Superpotentials for 
vector bundle moduli,''
Nucl.\ Phys.\ B {\bf 653} (2003) 400
[arXiv:hep-th/0205190].

\bibitem{Buchbinder:2002pr}
E.~I.~Buchbinder, R.~Donagi and B.~A.~Ovrut,
``Vector bundle moduli 
superpotentials in heterotic superstrings and M-theory,''
JHEP {\bf 0207} (2002) 066 
[arXiv:hep-th/0206203].

\bibitem{Buchbinder:wz}
E.~Buchbinder and B.~A.~Ovrut,
``Vector Bundle Moduli,
''
Russ.\ Phys.\ J.\  {\bf 45} (2002) 662
[Izv.\ Vuz.\ Fiz.\  {\bf 2002N7} (2002) 15].

\bibitem{Khoury:2001wf}
J.~Khoury, B.~A.~Ovrut, P.~J.~Steinhardt and N.~Turok,
``The ekpyrotic universe: Colliding branes and the origin of the hot big  bang,''
[hep-th/0103239].

\bibitem{Bastero-Gil:2002hs}
M.~Bastero-Gil, E.~J.~Copeland, J.~Gray, A.~Lukas and M.~Plumacher, 
``Baryogenesis by brane-collision,''
Phys.\ Rev.\ D {\bf 66} (2002) 066005
[arXiv:hep-th/0201040].

\bibitem{Lukas:1998hk}
A.~Lukas, B.~A.~Ovrut and D.~Waldram,
``Non-standard embedding and 
five-branes in heterotic M-theory,''
Phys.\ Rev.\ D {\bf 59} (1999) 106005
[arXiv:hep-th/9808101].

\bibitem{Derendinger:2001gy}
J.~Derendinger and R.~Sauser,
``A five-brane modulus in the 
effective N = 1 supergravity of M-theory,''
Nucl.\ Phys.\ B {\bf 598} (2001) 87
[hep-th/0009054].

\bibitem{Copeland:2001zp}
E.~J.~Copeland, J.~Gray and A.~Lukas,
``Moving five-branes in 
low-energy heterotic M-theory,''
Phys.\ Rev.\ D {\bf 64} (2001) 126003
[arXiv:hep-th/0106285].

\bibitem{Copeland:2002fv}
E.~J.~Copeland, J.~Gray, A.~Lukas and D.~Skinner,
``Five-dimensional 
moving brane solutions with four-dimensional limiting  behaviour,''
Phys.\ Rev.\ D {\bf 66} (2002) 
124007
[arXiv:hep-th/0207281].

\bibitem{Witten:1996gx}
E.~Witten,
``Small Instantons in String Theory,
''
Nucl.\ Phys.\ B {\bf 460
} (1996) 541
[hep-th/9511030].



\bibitem{Ganor:1996mu}
O.~J.~Ganor and A.~Hanany,
``Small $E8$ Instantons and Tensionless 
Non-critical Strings,''
Nucl.\ Phys.\ B {\bf 474} (1996) 122
[hep-th/9602120].

\bibitem{Strominger:et}
A.~Strominger,
``Heterotic Solitons,''
Nucl.\ Phys.\ B {\bf 343} (1990) 
167 [Erratum-ibid.\ B {\bf 353} (1991) 565].

\bibitem{Johnson:gi}
C.~V.~Johnson,
``D-Branes,''
{\it  Cambridge, USA: Univ. Pr. (2003) 548 p}.

\bibitem{Lukas:1998ew}
A.~Lukas, B.~A.~Ovrut and D.~Waldram,
``The ten-dimensional effective action of strongly coupled heterotic  string theory,''
Nucl.\ Phys.\ B {\bf 540} (1999) 230
[arXiv:hep-th/9801087].

\bibitem{Lalak:1997ti}
Z.~Lalak, A.~Lukas and B.~A.~Ovrut,
``Soliton solutions of M-theory on an orbifold,''
Phys.\ Lett.\ B {\bf 425} (1998) 59
[arXiv:hep-th/9709214].

\bibitem{Callan:dj}
C.~G.~Callan, J.~A.~Harvey and A.~Strominger,
``World Sheet Approach To Heterotic Instantons And Solitons,''
Nucl.\ Phys.\ B {\bf 359} (1991) 611.

\bibitem{Callan:1991ky}
C.~G.~Callan, J.~A.~Harvey and A.~Strominger,
``Worldbrane actions for string solitons,''
Nucl.\ Phys.\ B {\bf 367} (1991) 60.

\bibitem{Duff:1994an}
M.~J.~Duff, R.~R.~Khuri and J.~X.~Lu,
``String solitons,''
Phys.\ Rept.\  {\bf 259} (1995) 213
[arXiv:hep-th/9412184].

\bibitem{Dorey:2002ik}
See,
N.~Dorey, T.~J.~Hollowood, V.~V.~Khoze and M.~P.~Mattis,
``The calculus of many instantons,''
Phys.\ Rept.\  {\bf 371} (2002) 231
[arXiv:hep-th/0206063], and references therein.

\bibitem{Strominger:uh}
A.~Strominger,
``Superstrings With Torsion,''
Nucl.\ Phys.\ B {\bf 274} (1986) 253.

\bibitem{Bailin:nk}
D.~Bailin and A.~Love,
``Orbifold Compactifications Of String Theory,''
Phys.\ Rept.\  {\bf 315} (1999) 285.

\bibitem{Polchinski:rr}
J.~Polchinski,
``String Theory. Vol. 2: Superstring Theory And Beyond,''
{\it  Cambridge, UK: Univ. Pr. (1998) 531 p}.

\bibitem{jgal}
James Gray and Andr\'{e} Lukas - In preparation.

\bibitem{Lima:1999dn}
E.~Lima, 
H.~Lu, B.~A.~Ovrut and C.~N.~Pope,
``Instanton moduli and brane creation,
  ''
Nucl.\ Phys.\ B {\bf 569} (2000) 247
[arXiv:hep-th/9903001].
\end{thebibliography}
\end{document}